\documentclass[a4paper]{jpconf}
\usepackage{graphicx}
\usepackage{hyperref}
\usepackage{amsmath}
\usepackage{xfrac}

\newcommand{\sNN}{s_\mathrm{NN}}

\begin{document}
\title{Vorticity in the QGP liquid and hyperon polarization at the RHIC BES energies}

\author{Iurii Karpenko$^{1,2}$ and Francesco Becattini$^{1,3}$}

\address{$^1$ INFN - Sezione di Firenze, Via G. Sansone 1, I-50019 Sesto Fiorentino (Firenze), Italy}
\address{$^2$ Bogolyubov Institute for Theoretical Physics, 14-b, Metrolohichna st., 03680 Kiev, Ukraine}
\address{$^3$ Universit\'a di Firenze, Via G. Sansone 1, I-50019 Sesto Fiorentino (Firenze), Italy}

\ead{yu.karpenko@gmail.com}

\begin{abstract}
We calculate the polarization of $\Lambda$ hyperons in Au-Au collisions at RHIC Beam Energy Scan range $\sqrt{\sNN}=7.7, \dots, 200$~GeV in a state-of-the-art 3+1 dimensional cascade + viscous hydro model vHLLE+UrQMD. We find that the polarization of $\Lambda$ in the out-of-plane direction decreases substantially with collision energy. We explore the connection between the polarization signal and thermal vorticity and discuss the feed-down and hadronic rescattering effects on the mean polarization of all produced $\Lambda$ hyperons.
\end{abstract}

\section{Introduction}

Signatures of the Quark-Gluon Plasma (QGP) formation have been extensively studied via momentum space distributions of produced hadrons, but it is also natural to assume that finite angular momentum of the QGP droplets should be translated to alignment of final hadron's spins. Hydrodynamical approach is a well-established tool for the QGP studies, however, polarization of produced hadrons had not been systematically assessed in it.
In particular, the widely used Cooper-Frye formula assumes no distinction between different spin projections of the produced hadrons. Spin only enters in the distribution function part of the Cooper-Frye formula as a $(2J+1)$ multiplier.
However, it was shown in \cite{Becattini:2013fla} that a more subtle effect is present and can play a role: spin-vorticity coupling can induce non-vanishing polarization of hadrons with nonzero spin produced in local thermal equilibrium. This is a purely thermo-mechanical effect, which polarizes both particles and anti-particles in the same direction defined by local thermal vorticity of the medium.

Up to now, few results for the polarization of $\Lambda$ baryons based on this effect have been reported. In \cite{Becattini:2013vja}, simulation of noncentral $\sqrt{s_{\rm NN}}=200$~GeV Au-Au collisions at RHIC with initial state from Yang-Mills dynamics followed by a 3~dimensional ideal hydro expansion resulted in polarization of low-$p_T$ $\Lambda$ of few percent magnitude, parallel to the direction of total angular momentum of the fireball, whereas for few GeV $p_T$ it reaches 9\%. In another work \cite{Becattini:2015ska} aiming at the same collision energy, a 3 dimensional hydrodynamic expansion with initial state from optical Glauber model with parametrized rapidity dependence chosen to reproduce directed flow results \cite{Bozek:2010bi} resulted in much lower values of polarization: about 0.2\% for low-$p_T$ $\Lambda$ and up to 1.5\% for $z$ component of polarization of high-$p_T$ $\Lambda$. In a more recent work \cite{Pang:2016igs} an event-by-event 3 dimensional viscous hydrodynamics with initial state from AMPT model results in similar few per mille average polarization for A+A collisions at $\sqrt{\sNN}=62.4, 200$ and $2760$~GeV.

The abovementioned results confront the experimental analysis of Au-Au collisions at $\sqrt{s_{\rm NN}}=200$~GeV by STAR \cite{Abelev:2007zk} which only puts an upper limit of 0.02 on the polarization of $\Lambda$ baryons. At lower collision energies, because of emerging baryon stopping effect one may expect larger local vorticity and, as a result, larger polarization signal. Indeed, preliminary STAR results \cite{MLisaTalk} show that the mean polarization of hyperons can reach as much as several percents in Au-Au collisions at lower RHIC Beam Energy Scan (BES) energies. In this report we give an assessment of the polarization signal in a state-of-the-art hybrid model, which is tuned to reproduce basic hadron observables for heavy ion collisions in the BES program: (pseudo)rapidity, transverse momentum distributions and elliptic flow coefficients.

\section{Polarization observable in the cascade+viscous hydrodynamic model}

We simulate heavy ion collisions at RHIC Beam Energy Scan energies in \texttt{vHLLE+UrQMD} hybrid \cite{Karpenko:2015xea}. Herein we summarize its main features. The initial state is modeled with \texttt{UrQMD} cascade \cite{Bass:1998ca, Bleicher:1999xi}. At a hypersurface $\tau=\sqrt{t^2-z^2}=\tau_0$ the fluidization is imposed. At lower BES energies $\tau_0$ corresponds to a moment when two nuclei have completely passed through each other, $\tau_0=2R/(\gamma v_z)$. Therefore the duration of pre-hydro stage changes considerably, from 3.2~fm/c at $\sqrt{\sNN}=7.7$~GeV to 0.4~fm/c at $\sqrt{\sNN}=200$~GeV.
Following 3 dimensional hydrodynamic expansion is numerically solved with \texttt{vHLLE} code \cite{Karpenko:2013wva}. In the hydrodynamic phase, a finite effective value of shear viscosity over entropy density $\eta/s$ is taken into account, and the equation of state of the medium with finite $\mu_{\rm B}$ is based on Chiral model \cite{Steinheimer:2010ib}.

It was shown in \cite{Becattini:2013fla} that the mean relativistic spin vector of spin \sfrac{1}{2} particles of sort $i$ with four-momentum $p$, produced around point $x$ on particlization hypersurface is:
\begin{equation}\label{eq-Pixp}
 \Pi_i^\mu(x,p)=\frac{1}{8m_i} (1-f_i(x,p)) \epsilon^{\mu\nu\rho\sigma} p_\sigma \varpi_{\sigma\nu}
\end{equation}
where $f_i(x,p)$ is (Fermi-Dirac) distribution function of the particles and $\varpi_{\mu\nu} = -1/2 ( \partial_\mu \beta_\nu - \partial_\nu \beta_\mu)$ is thermal vorticity, equal to (minus) the antisymmetric part of the gradient of the four-temperature field $\beta^\mu = (1/T) u^\mu$, where $T$ is the proper temperature and $u^\mu$ is the hydrodynamic 4-velocity. The mean spin vector of the hadron species $i$, produced at particlization surface (which we will call \textit{direct} hadrons), can be then calculated using an extension of the Cooper-Frye formula:
\begin{align}\label{Piav}
 \langle \Pi_i^{*\mu} \rangle = \frac{1}{N_i} \int \frac{d^3p}{p^0} \int d\Sigma_\lambda p^\lambda f_i(x,p) \Pi_i^{*\mu}(x,p)
\end{align}
where $N_i=\int\frac{d^3p}{p^0}\int d\Sigma_\lambda p^\lambda f_i(x,p)$ is the average number of the hadrons produced, and Lorentz transformation is made to the rest frames of the particles, denoted with the asterisk. Finally we note that the experimentally observable quantity is mean polarization vector $P^\mu=\Pi^\mu/S$, where $S$ is the spin of particle.

\section{Results and conclusions}

\begin{figure}
\vspace*{-10pt}
\includegraphics[width=0.5\textwidth]{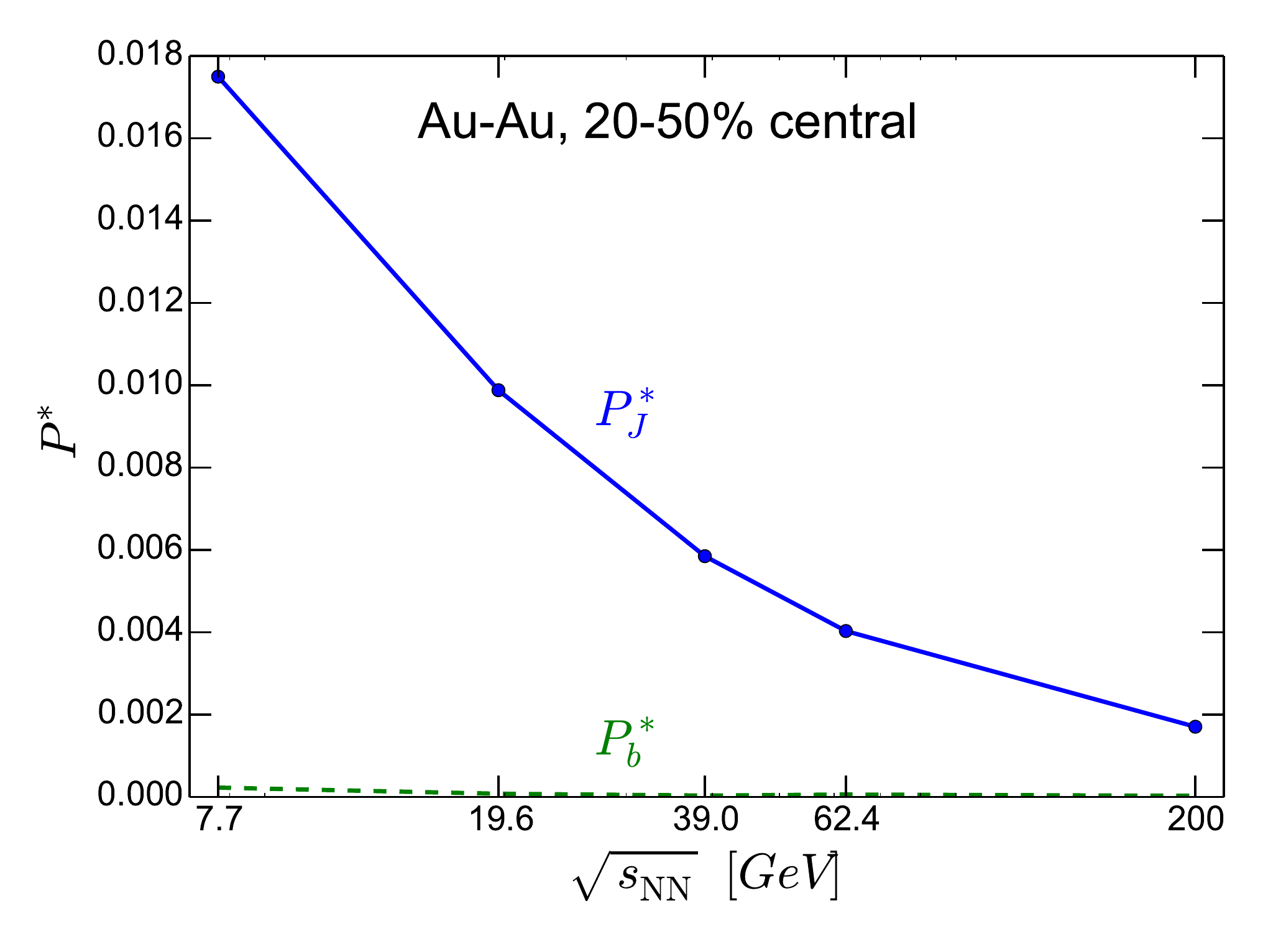}
\includegraphics[width=0.5\textwidth]{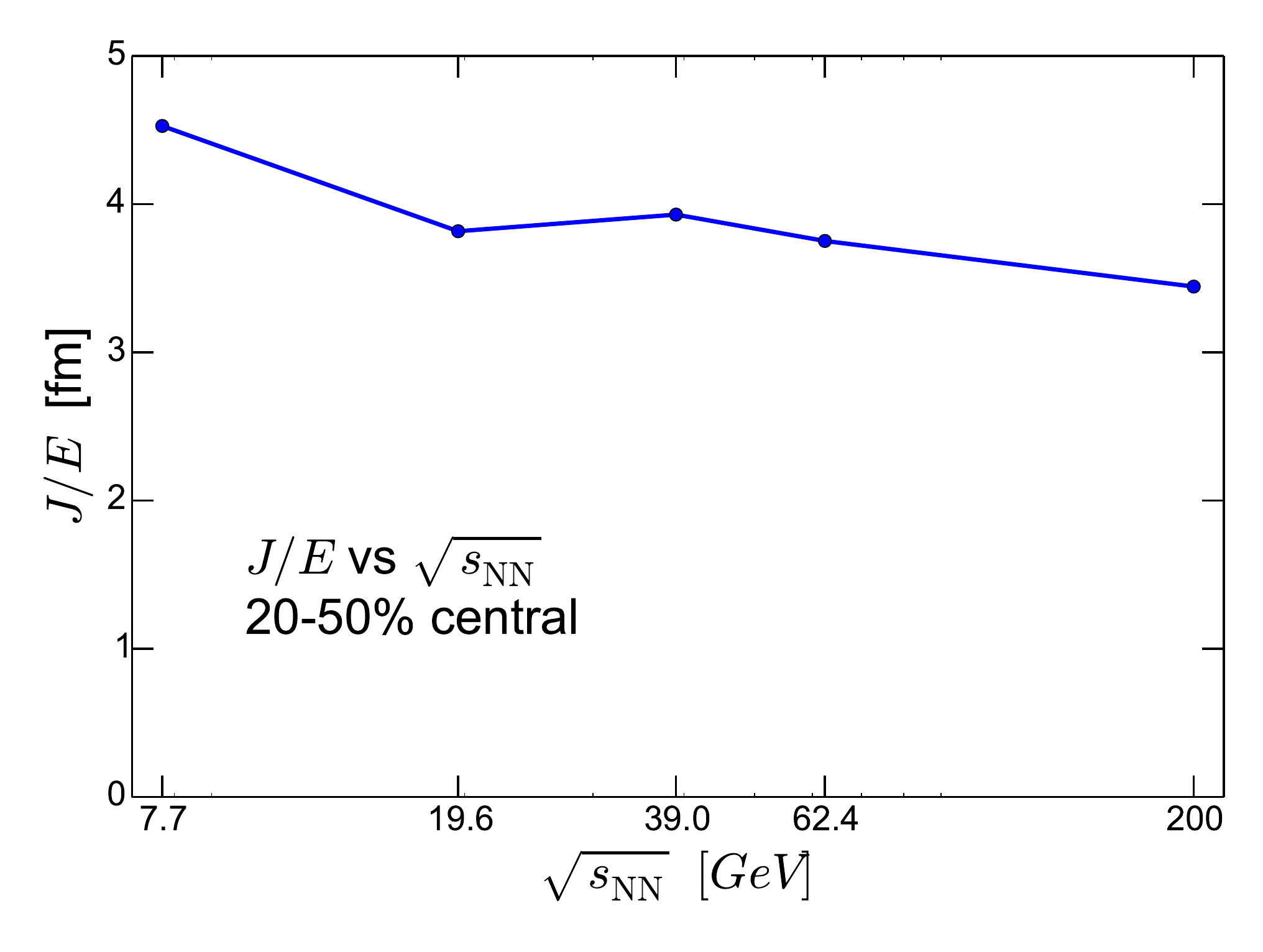}
\vspace*{-15pt}
\caption{Components of mean polarization vector of midrapidity $\Lambda$ produced at the particlization surface (left) and ratio of total angular momentum to total energy of the fireball (right), calculated in the model for 20-50\% central Au-Au collisions at $\sqrt{\sNN}=7.7,...,200$~GeV.}\label{fig1}
\end{figure}

\begin{figure}
\vspace*{-10pt}
\includegraphics[width=0.46\textwidth]{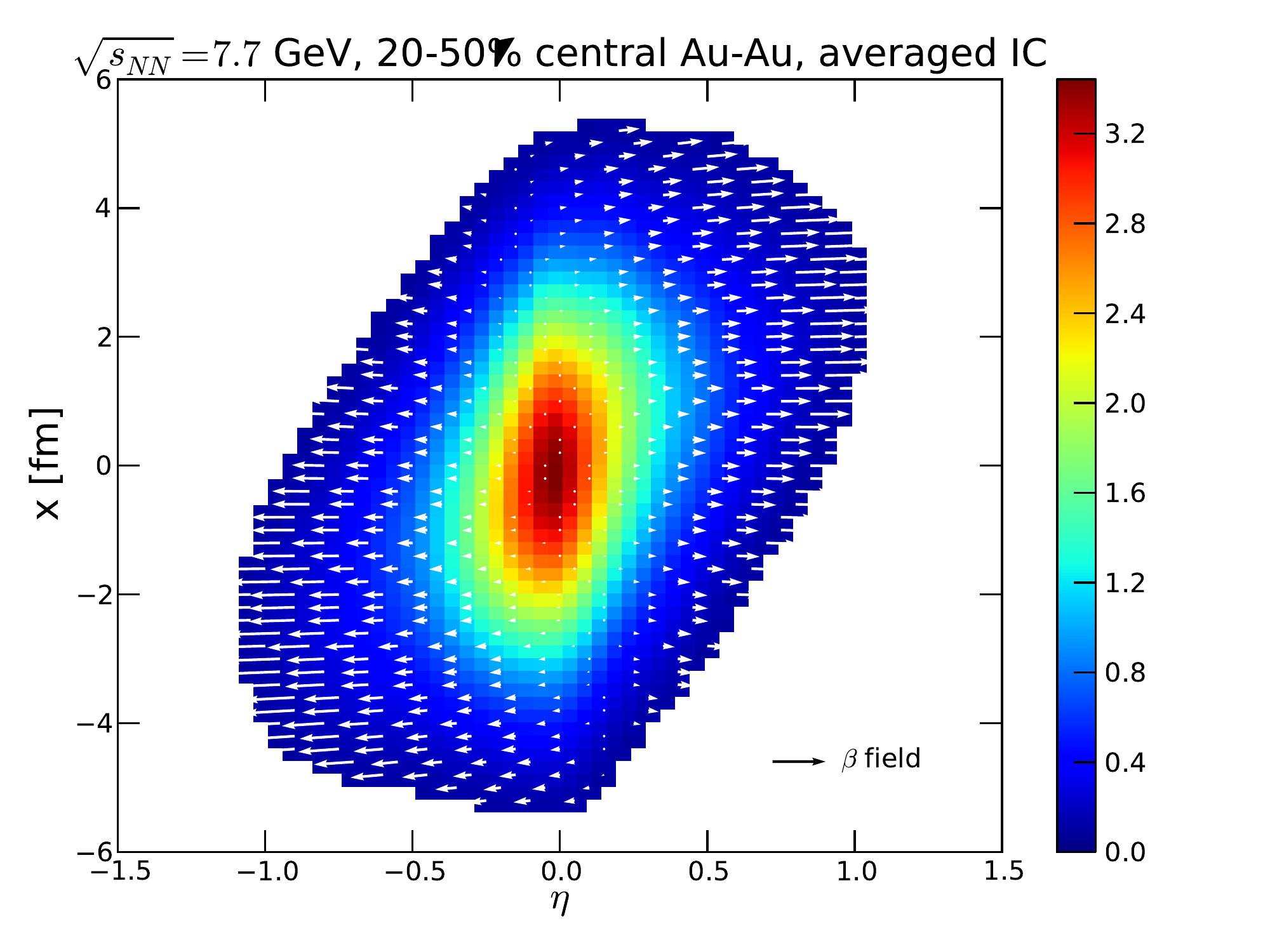}\hspace{-10pt}
\includegraphics[width=0.47\textwidth]{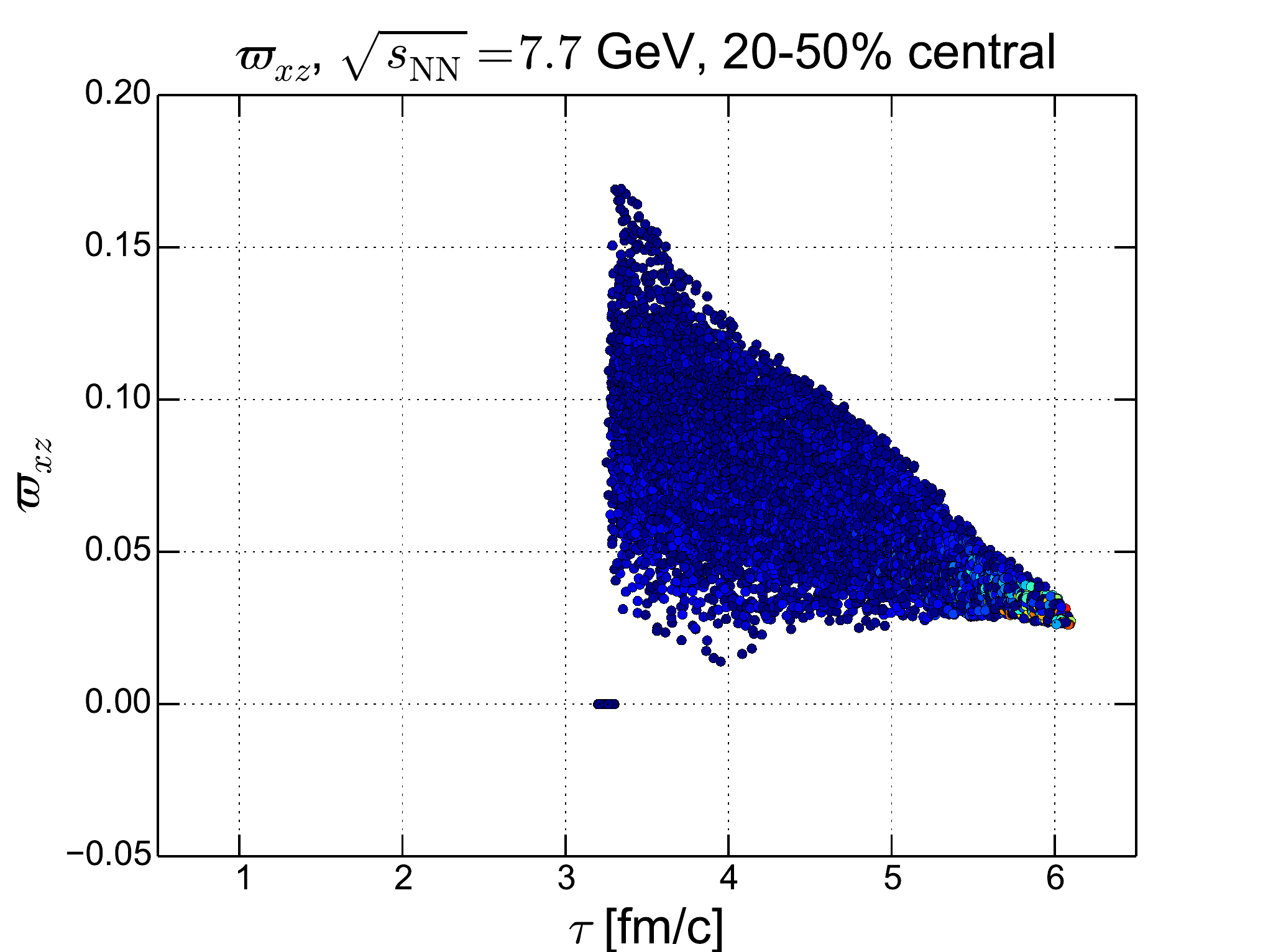}\\
\includegraphics[width=0.46\textwidth]{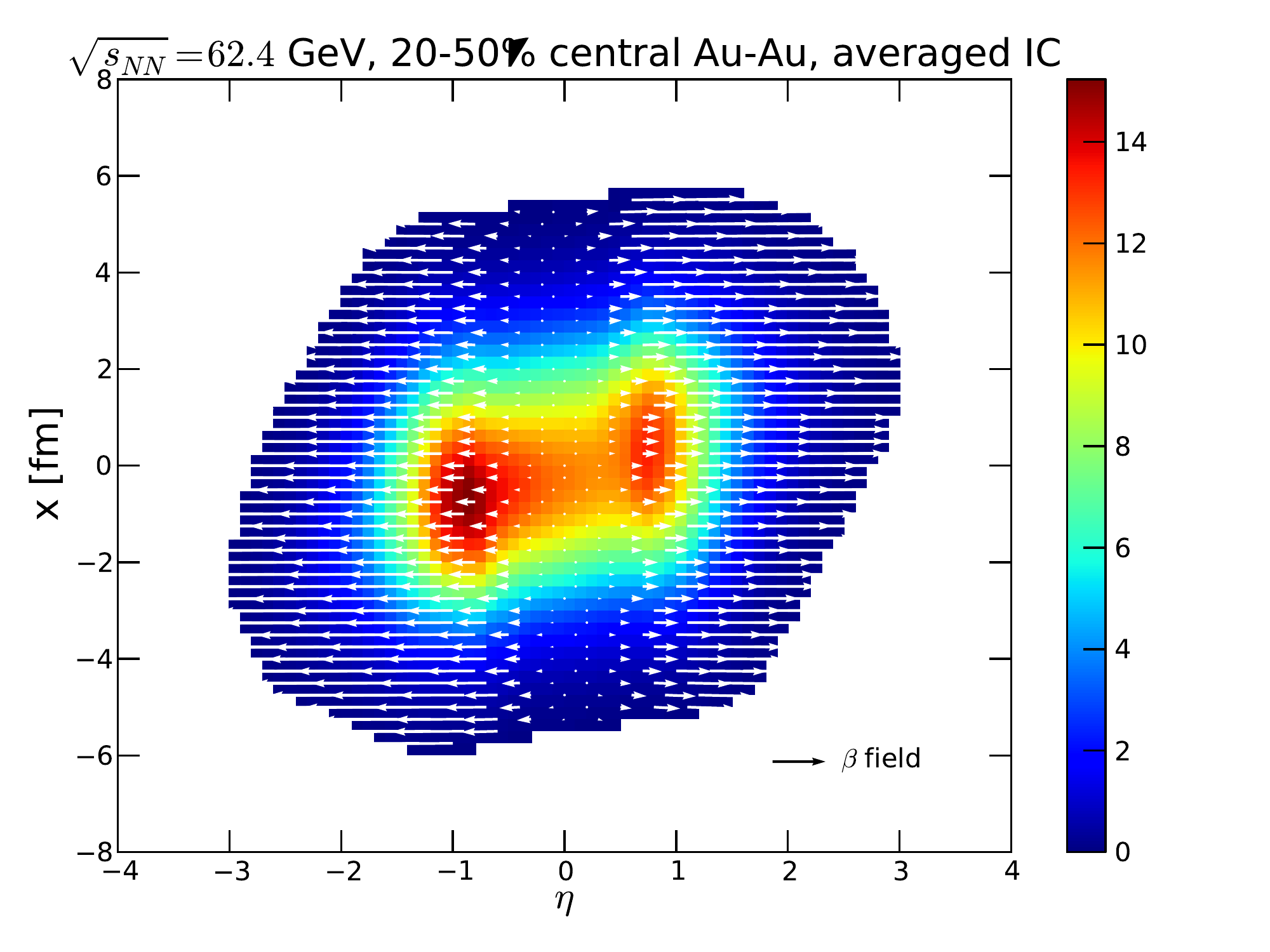}
\includegraphics[width=0.47\textwidth]{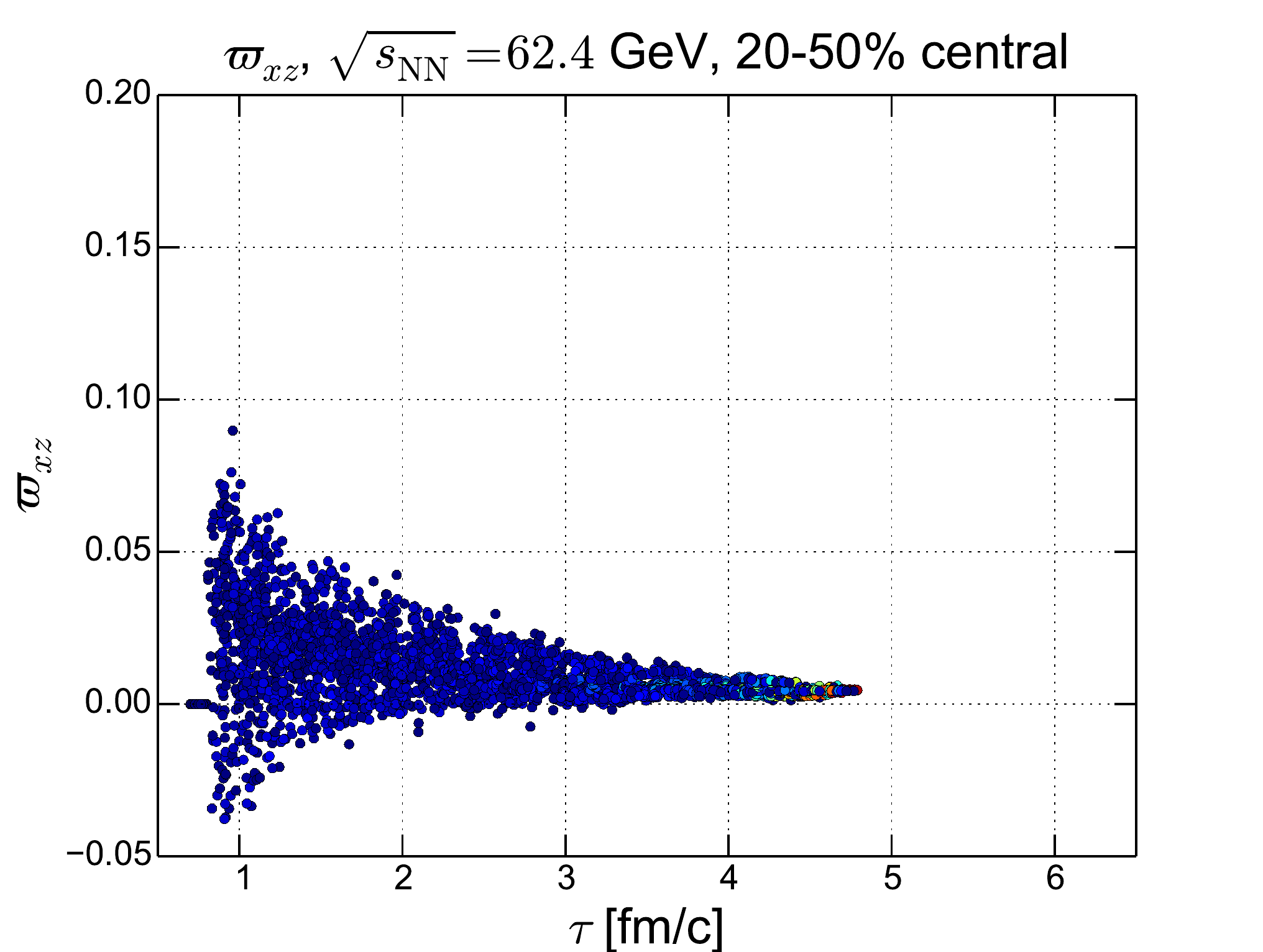}
\vspace*{-5pt}
\caption{Initial energy density profiles for hydrodynamic stage with arrows depicting initial beta field superimposed (left column) and $\varpi_{xz}$ over midrapidity ($|y|<0.3$) slice of particlization surface, projected onto time axis (right column). The hydrodynamic evolutions start from averaged initial states corresponding to 20-50\% central Au-Au collisions at $\sqrt{s_{\rm NN}}=7.7$ (top row) and $62.4$~GeV (bottom row).}\label{fig2}
\end{figure}

The resulting components of the mean polarization vector of direct $\Lambda$ baryons at midrapidity, calculated in the model for 20-50\% central Au-Au collisions at RHIC BES energies is presented on the left panel of Fig.~\ref{fig1} as a solid line. The $P_J$ is a component perpendicular to the reaction plane, and $P_b$ is the one parallel to the impact parameter. We observe a rapid decrease of the mean $\Lambda$ polarization with collision energy from a maximal value of about 1.8\% at $\sqrt{\sNN}=7.7$~GeV. While qualitatively one expects that the polarization is related to the angular momentum of the fireball, we argue that quantitatively it is not the case. In fact the total angular momentum of the fireball increases with increasing collision energy as one can see on the right panel of Fig.~\ref{fig2}, and the ratio of total angular momentum to total energy $J/E$ exhibits only a mild decrease.

As one can see from Eq.~\ref{eq-Pixp}, the actual quantity generating local polarization of hadrons is the local thermal vorticity at the points of their production. Having an access to the space-time flow structure in the model, we can explore how the vorticity changes with collision energy. On the right panel of fig.~\ref{fig2} we show a time evolution of the $xz$ component of thermal vorticity, which generates the dominant contribution to the out-of-plane component of polarization $P_J$, at two different collision energies. We find that (1) for lower collision energy, $\sqrt{\sNN}=7.7$~GeV the hydrodynamic expansion starts with higher value of $\varpi_{xz}$, which is a result of shear flow structure (plotted with white vectors on the left panel plot) at midrapidity formed by baryon stopping. Also, (2) shorter lifetime of hydro phase (as compared to the simulation at $\sqrt{\sNN}=62.4$~GeV) dilutes the initial vorticity less, resulting in larger vorticity at the end of hydrodynamic phase. As the low-$p_T$ $\Lambda$ are more likely to be emitted at the end of the hydrodynamic phase, the abovementioned factors explain the steep decrease of the mean $\Lambda$ polarization with collision energy. Similar trends - increase of angular momentum of the fireball and decrease of the mean vorticity with increasing collision energy - were observed in AMPT microscopic model \cite{Jiang:2016woz}.

Large fraction of observed $\Lambda$ come from decays of higher mass hyperon resonances, which possess polarization as well. The most abundant ones are strong decay of $\Sigma(1385)$ and electromagnetic decay of $\Sigma^0$. Preliminary analysis of the feed-down contributions and polarization transfer in the hyperon decays shows us that feed-down contributions from $\Sigma^0$ and $\Sigma(1385)$ decrease the resulting mean polarization by only few percents. However, extending the list of decays up to $\Sigma(1670)$ results in about 15\% suppression of the overall $\Lambda$ polarization, as $\Lambda$ from decays of heavier resonances acquire alternating signs of polarization \cite{Becattini:2016gvu}.

In addition, in UrQMD cascade which is used to treat interactions after particlization, cross sections of $\Lambda$ and $\Sigma^0$ with mostly abundant mesons and baryons, which are calculated according to Additive Quark Model, are comparable to those of nucleons \cite{Bleicher:1999xi}. This implies that $\Lambda$ actively rescatter in the hadronic phase. Scatterings will presumably randomize the spin direction of primary as well as secondary particles, thus decreasing the estimated mean global polarization. Therefore the values of polarization on Fig.~\ref{fig1} should serve only as an upper bound for the polarization of all $\Lambda$ in a complete hydro+cascade model with given initial conditions.

\section{Acknowledgements}
This work was partly supported by the University of Florence grant \textit{Fisica dei plasmi relativistici: teoria e applicazioni moderne}.

\end{document}